\documentclass[preprint,12pt]{elsarticle}

\usepackage{lscape}

\usepackage{graphicx}

\usepackage{amssymb}
\usepackage[mathlines]{lineno}

\begin{document}

\begin{frontmatter}

\title{$N$-body simulations of oligarchic growth of Mars: Implications for Hf-W chronology }

\author[ucla,jpl]{Ryuji Morishima\corref{cor1}}
\ead{Ryuji.Morishima@jpl.nasa.gov}

\author[ens,eth]{Gregor J. Golabek}

\author[bay,cnrs]{Henri Samuel}

\cortext[cor1]{Corresponding author}

\address[ucla]{University of California, Los Angeles, Institute of Geophysics and Planetary Physics, Los Angeles, CA  90095, USA}

\address[jpl]{Jet Propulsion Laboratory/California Institute of Technology, Pasadena, CA 91109, USA}

\address[ens]{ENS de Lyon, Laboratoire de G\'{e}ologie de Lyon, Lyon 69364, France}

\address[eth]{Institute of Geophysics, ETH Zurich, Sonneggstrasse 5, 8092 Z\"{u}rich, Switzerland}

\address[bay]{Bayerisches Geoinstitut, Universit\"{a}t Bayreuth, 95447 Bayreuth, Germany}

\address[cnrs]{Institut de Recherche en Astrophysique et PlanŽtologie, CNRS, 31400, Toulouse, France}

\begin{abstract}
Dauphas and Pourmand (2011) [Nature 473, 489--492] estimated the accretion timescale of Mars to be 
1.8 $^{+0.9}_{-1.0}$ Myr from the W isotopes of martian meteorites. 
This timescale was derived assuming perfect metal-silicate equilibration between the impactor and the target's mantle. 
However, in the case of a small impactor most likely only a fraction of the target's mantle is involved in the equilibration, 
while only a small part of the impactor's core equilibrates in the case of a giant impact.
We examined the effects of imperfect equilibration
using results of high-resolution $N$-body simulations for the oligarchic growth stage.   
These effects were found to be small as long as a planetary embryo has a deep liquid magma ocean during its accretion.
The effect due to partial involvement of the target's mantle in equilibration is small due to the low metal-silicate partition coefficient for W suggested from 
the low Hf/W ratio of the martian mantle. 
The effect due to partial involvement of the impactor's core is also small because a large fraction of the embryo mass is delivered from small planetesimals, 
which are likely to fully equilibrate in the deep magma ocean on the embryo. 
The accretion timescale of Mars estimated by the Hf-W chronology is shorter than that expected for the minimum mass solar nebula model
as long as more than 10\% of each impactor's core re-equilibrates with the martian mantle and the final stages of accretion are prolonged. 
This probably indicates that accretion of Mars proceeded rapidly due to solid and gas surface 
densities significantly larger than those for the minimum mass solar nebula or due to accretion of small fragments or pebbles.
\end{abstract}


\end{frontmatter}


\section{Introduction}
The theory of planet formation suggests that tens of Mars-sized embryos form from planetesimals in the inner solar system during the runaway and oligarchic 
growth stages, which last for  $\sim$ 0.1-10 Myr (Kokubo and Ida, 1998; Wetherill and Steward 1993).
In the subsequent,  much longer ($\sim 100$ Myr) giant-impact stage,  mutual collisions between embryos occur (Morishima et al., 2010; O'Brien et al., 2006). 
Mars is considered likely to be a remnant embryo because its mass 
is close to the theoretically predicted final mass of an oligarchic embryo, commonly referred to as the isolation mass (Lissauer et al., 1987),
and its accretion timescale suggested from the Hf-W chronology (Dauphas and Pourmand, 2011; Nimmo and Kleine, 2007)
is much shorter than that for the Earth-Moon system (K\"{o}nig et al., 2011; Touboul et al., 2007).
The giant impact stage is rather stochastic, because the timing of the final large impact has a large dispersion 
in $N$-body simulations starting with similar initial conditions (Morishima et al., 2010; O'Brien et al., 2006).
However, the oligarchic growth stage is deterministic and the mass evolution of an embryo can even be expressed 
as a simple analytic solution if a uniform size of planetesimals is adopted (Chambers, 2006). 
Therefore,  if the accretion timescale of Mars is precisely determined by the Hf-W chronology,
several important quantities which determine the accretion timescale may be retrieved,  
such as the gas and solid surface densities of the protosolar disk, and the planetesimal size.

Dauphas and Pourmand (2011) estimated the accretion timescale of Mars, most precisely to date, to be 1.8 $^{+0.9}_{-1.0}$ Myr. 
This timescale was derived using the isotopic evolution model of Jacobsen (2005) which assumes 
perfect metal-silicate equilibration between the impactor and the target's mantle during an impact 
(the more massive body is designated to be the target, while the less massive one is the impactor). 
However, two possible types of imperfect equilibration are conceivable; 
both of them result in a longer accretion timescale.

The first one is that only a part of the metallic core of the impactor is involved in equilibration.
Mezger et al. (2012) showed that if $\sim$ 40 \% of each impactor's core re-equilibrates with the martian mantle,
 the estimated accretion time scale of Mars becomes a few times longer than in the case of perfect equilibration.
Nimmo et al. (2010) investigated tungsten isotopic evolution of terrestrial planets using 
outputs of $N$-body simulations of O'Brien et al. (2006) and found 
that in order to produce the terrestrial tungsten anomaly, the degree of equilibration is required to be 30-80 \%. 
Rudge et al. (2010) adopted various growth curves of the Earth and found that both perfect equilibrium and partial equilibrium model
can reproduce the terrestrial tungsten anomaly and the abundances of siderophile elements.  
They also constrained the degree of equilibration to be at least 36\%.

Perfect equilibration requires emulsification of the impactor's core down to cm-scale droplets 
in the target's mantle (Rubie et al., 2003) because core formation is likely to occur even for a small impactor (down to $\sim$ 10 km)
due to heating by radiogenic decay of $^{26}$Al and $^{60}$Fe (Moskovitz and Gaidos, 2011; Neumann et al., 2012).    
Whether sufficient emulsification occurs when a metallic core sinks in a liquid magma ocean 
was investigated using different types of hydrodynamical models 
(Dahl and Stevenson, 2010; Deguen et al., 2011; Ichikawa et al., 2010; Samuel, 2012). 
Except for Dahl and Stevenson (2010), these studies consistently show that deformation and breakup of the iron core
(neglected in Dahl and Stevenson's analytic model)
is the main mechanism that leads to very efficient emulsification and metal-silicate equilibration,
as long as the core size is smaller than the depth of the magma ocean on the target.
Kendall and Melosh (2012) showed that significant emulsification of the impactor's core 
already occurs during a high-velocity impact (they adopted an impact speed of 11.5 km/s).

The second one, which has been paid less attention than the first one, 
is that only a part of the target's mantle is involved in equilibration.
This is likely to occur if the impactor size is much smaller than the target size; 
such a situation is common during the oligarchic growth stage. 
Even if the impactor size is large,  an upper metal-rich layer 
and a lower metal-poor layer may overturn due to the Rayleigh-Taylor instability, 
which may reduce the volume fraction of the target's mantle involved in equilibration (Sasaki and Abe, 2007).

As discussed above, the impactor-to-target mass ratio is likely to be 
the important parameter for metal-silicate equilibration. 
The mass distribution of impactors during the oligarchic growth stage is not clearly known.
During the former runaway growth stage, the mass distribution can be described by a power-law
with the exponent of $\sim - 2.5$ (Kokubo and Ida, 2000; Ormel et al., 2010). 
During the oligarchic growth stage, large embryos separate from the continuous 
size distribution.  In addition, mutual collisions between oligarchic bodies 
are expected. Chambers (2006) estimated that one-third of the mass of 
an oligarchic embryo is delivered by 
embryo-embryo collisions, provided that the mutual separation normalized 
by the Hill radius is fixed. This needs to be examined by direct $N$-body simulations.

In the present paper, we investigate accretion timescales of embryos and 
mass distribution of impactors, by conducting high resolution $N$-body simulations
of the oligarchic growth stage. 
Some basics of the oligarchic growth stage are reviewed in Sec.~2.
In Sec.~3, methods and results of two $N$-body simulations are shown; 
one case  with nebular gas and another case without gas.
In Sec.~4, using the results of $N$-body simulations, 
we model the isotopic evolution of the Hf-W system and examine the effects 
of possible imperfect  metal-silicate equilibration. 
Discussion and summary are given in Sec.~5 and Sec.~6.

\section{Expected time evolution of the mass of an oligarchic body} 
Before discussing the results of $N$-body simulations, some basics of the oligarchic 
growth stage are briefly reviewed.
Consider planetary embryos surrounded by small planetesimals.
An embryo gravitationally influences planetesimals in an annulus around the embryo's orbit 
and these nearby planetesimals collide with the embryo.
This annulus is called the feeding zone and its radial width is roughly equivalent to the 
orbital separation between neighboring embryos and is 
known to become $b_{\rm e}r_{\rm H}$ as a result of orbital repulsion between embryos (Kokubo and Ida 1998),
where $b_{\rm e} \sim 10$ is the scaling factor and $r_{\rm H}$ is the Hill radius
of the embryo given by
\begin{equation}
r_{\rm H} = a\left(\frac{2m_{\rm e}}{3M_{\odot}}\right)^{1/3} = 2^{1/3}ah,
\end{equation}
where $a$ is the semimajor axis of the embryo, $m_{\rm e}$ is the mass of the embryo,
$M_{\odot}$ is the solar mass, and $h$ is the reduced Hill radius used below. 

Defining the mass fraction of an embryo to the total solid mass in its feeding zone to be $f_{\rm e}$,
the mass of the embryo  $m_{\rm e}$ is given by 
\begin{eqnarray}
m_{\rm e} &=& 2\pi f_{\rm e} a b_{\rm e} r_{\rm H} \Sigma_{\rm solid}  = (2\pi f_{\rm e} b_{\rm e}\Sigma_{\rm solid})^{3/2} a^3 \left(\frac{2}{3M_{\odot}}\right)^{1/2} \nonumber \\
&=& 1.8 \left(\frac{f_{\rm e}}{1}\right)^{3/2} \left(\frac{b_{\rm e}}{10}\right)^{3/2} \left(\frac{\Sigma_{\rm solid}}{5.0 \hspace{0.3em} {\rm g \hspace{0.2em} cm}^{-2}}\right)^{3/2}
\left(\frac{a}{1.5 \hspace{0.3em}{\rm AU}}\right)^{3/2}m_{\rm Mars}, \label{eq:mp}
\end{eqnarray}
$\Sigma_{\rm solid}$ is the solid surface density and $m_{\rm Mars}$ is the mass of Mars (6.4185 $\times 10^{26}$ g). 
When an embryo sweeps up all solid material in its feeding zone  ($f_{\rm e} = 1$), the embryo mass is called the isolation mass (Lissauer et al., 1987):
\begin{equation}
m_{\rm iso} = m_{\rm e} (f_{\rm e} = 1).
\end{equation}

The time evolution  of $f_{\rm e}$ is given by (Chambers, 2006)
\begin{equation}
\frac{d f_{\rm e}}{dt} = A f_{\rm e}^{1/2}(1-f_{\rm e}), \label{eq:dsige}
\end{equation}
where $A$ is
\begin{equation}
A = \frac{31.7 C}{b_{\rm c}^{1/2}\tilde{e}^2}\left(\frac{\Sigma_{\rm solid}^{1/2}}{P\rho^{1/3}M_{\odot}^{1/6}}\right), \label{eq:aa}
\end{equation}
where $b_{\rm c} = 2^{1/3}b_{\rm e}$, $\tilde{e}$ is the orbital eccentricity of planetesimals normalized by $h$, $P$ is the orbital period, 
$\rho$ is the density of nebular gas.
The factor C in Eq.~(\ref{eq:aa}) represents acceleration of growth due to embryo-embryo collisions 
and we set $C = 1.5$ as estimated in Chambers (2006).
If viscous stirring of embryos and gas drag are in equilibrium and  the radius $r$ of planetesimals is uniform,  
$\tilde{e}$ is written as
\begin{equation}
\tilde{e} = 2.7\left(\frac{r\rho}{b_{\rm c}C_{\rm D}a\rho_{\rm gas}}\right)^{1/5}, \label{eq:tile}
\end{equation}
where $C_{\rm D}$ is the drag coefficient assumed to be unity and $\rho_{\rm gas}$ is the gas density.
Eqs.~(\ref{eq:aa}) and (\ref{eq:tile}) mean that planetesimals frequently 
collide with an embryo, if $\tilde{e}$ is low due to a large $\rho_{\rm gas}$ or a small $r$ 
or simply if $\Sigma_{\rm solid}$ is large.

The solution of Eq.~(\ref{eq:dsige}) is 
\begin{equation}
f_{\rm e}(t)^{1/2} = \tanh\left[\int_{0}^{t} \frac{A(t')}{2} dt'  + {\rm atanh}(f_{\rm e}(0)^{1/2}) \right],  \label{eq:sige}
\end{equation}
and the time evolution of the embryo's mass is given as
\begin{equation}
m_{\rm e}(t) = m_{\rm iso}f_{\rm e}(t)^{3/2}.  \label{eq:me}
\end{equation}
If $A$ is independent of time and $f_{\rm e}(0)$ is negligible, Eq.~(\ref{eq:me}) reduces to 
\begin{equation}
m_{\rm e}(t)= m_{\rm iso}\tanh^3\left(\frac{t}{\tau_{\rm grow}} \right), \label{eq:me2}
\end{equation}
where 
\begin{equation}
\tau_{\rm grow} = \frac{2}{A}. 
\end{equation}
Eq.~(\ref{eq:me2}) shows that the embryo gains 44\%, 90\%, and 99\% of its final mass during timescales $1 \tau_{\rm grow}$, $2 \tau_{\rm grow}$, and $3 \tau_{\rm grow}$. 
The timescale $\tau_{\rm grow}$ for Mars was estimated by Dauphas and Pourmand (2011)
as 1.8 $^{+0.9}_{-1.0}$ Myr  by setting $m_{\rm iso}$ to be the mass of Mars. 
The parameter $A$ is usually time-dependent because gas dissipates with time. 
Even in such a case, an analytic expression can be obtained as long as $A$ is analytically integrable with respect to time; 
for example, in the case where gas dissipates exponentially with time.

To estimate $f_{\rm e}(0)$,  the embryo's mass at the transition from the runaway growth stage 
to the oligarchic growth stage is discussed. 
If the mass distribution is given by  $dn = km^{q}dm$ (where $n$ is the cumulative number of bodies inside the feeding zone of the embryo and $k$ is a constant),
$q$ is $\simeq -2.5$ at this transition 
(Kokubo and Ida, 2000; Ormel et al., 2010).
The total mass  inside the feeding zone is given by
\begin{equation}
m_{\rm T} = \int_{m_{\rm 0}}^{m_{\rm e}}m \ dn = \frac{q+2}{q+1} \left[\left(\frac{m_{0}}{m_{\rm e}}\right)^{-q-2}-1\right]m_{\rm e}, \label{eq:mt}
\end{equation}
where $m_0$ is the mass of the smallest planetesimal.
In Eq.~(\ref{eq:mt}), we removed $k$ using the definition of the largest body (embryo) given by 
\begin{equation}
\int_{m_{\rm e}}^{\infty} \ dn = -\frac{km_{\rm e}^{q+1}}{q+1} = 1. 
\end{equation}
With $m_{\rm e} \gg m_{0}$ and $q = -2.5$, the mass fraction of the embryo relative to the total mass is 
\begin{equation}
f_{\rm tr} = \frac{m_{\rm e}}{m_{\rm T}} = \frac{1}{3}\left(\frac{m_{0}}{m_{\rm e}}\right)^{1/2}. \label{eq:ftr}
\end{equation}
Inserting Eq.~(\ref{eq:ftr}) into Eq.~(\ref{eq:mp}), the transition mass is given by 
\begin{equation}
m_{\rm tr} = m_{\rm e} (f_{\rm e} = f_{\rm tr}) = \left(\frac{1}{3}\right)^{6/7}m_{\rm iso}^{4/7}m_0^{3/7}.  \label{eq:mtr}
\end{equation}
This mass coincides within a factor of two with  the transition mass derived from the numerical simulations of Ormel et al. (2010) (their Eq.~(13)).
 We adopt  $f_{\rm e}(0) = (m_{\rm tr}/m_{\rm iso})^{2/3}$ from Eq.~(\ref{eq:me}).
 
\section{$N$-body simulations}
\subsection{Methods}
Two $N$-body simulations were performed for systems in annuli around the current location of Mars, 1.5 AU; 
one with nebular gas (Sim.~A) and the second one without gas (Sim.~B). 
All input parameters are the same for both simulations except those for nebular gas.
Each impact between two bodies is assumed to result in perfect merging. 
To keep the total surface density of solid bodies in a simulation annulus constant,  the body supply boundary condition (Kokubo and Ida, 2000)
is adopted; if the semimajor axis $a$ of a planetesimal is larger than the outer boundary, $a_{\rm out}$,  
this planetesimal is sent to the inner boundary, $a_{\rm in}$, without changing its orbital eccentricity $e$ and inclination $i$.
With this boundary condition, planetesimals near the boundaries tend to confine embryos inside the annulus, 
mimicking viscous stirring of embryos outside the boundaries (Kokubo and Ida 2000). 

The total mass $M_{\rm d}$ of planetesimals in the annulus is $2.0 \times 10^{27}$ g ($\simeq 3 m_{\rm Mars}$).  
The initial surface density of planetesimals is set to  
\begin{equation}
\Sigma_{\rm solid}(a) = \left(\frac{a}{\rm 1.5 \hspace{0.3em} AU}\right)^{-3/2} \Sigma_{\rm solid}({\rm 1.5 \hspace{0.3em}AU}),
\end{equation}
where  $\Sigma_{\rm solid}({\rm 1.5 \hspace{0.3em}AU})$ is adopted to be 5.0 g cm$^{-2}$.
This is higher than that for the Minimum Mass Solar Nebula (MMSN; Hayashi, 1981) by 30\%.
The width of the simulation annulus $\Delta a$ derived using the above  $M_{\rm d}$ and $\Sigma_{\rm solid}$ is 0.1896 AU.
The boundary radii are then given as $a_{\rm in} = 1.5$ AU $- \Delta a/2$ and $a_{\rm out} = 1.5$ AU $+ \Delta a/2$.
The initial number of planetesimals is 5,000 and the physical density $\rho$ is $3.95$ g cm$^{-3}$ for all bodies.
The initial mass distribution of planetesimals is given by a single power-law with $q = -2.5$.
The initial mass ratio between the largest and the smallest bodies is set to be 20. This gives the mass of 
the smallest planetesimal $m_0 = 1.7 \times$ $10^{23}$ g.

For the simulation with gas,
the gas-to-solid ratio averaged over the annulus is set to be 240 at the beginning, as for the MMSN.
The gas temperature is given as  $T= 220(r/1.5 {\rm AU})^{-1/2}$ K.
The surface density of gas, $\Sigma_{\rm gas}$, is assumed to decay exponentially with a timescale of 2 Myr:
\begin{equation}
\Sigma_{\rm gas}(a,t) = \left(\frac{a}{1.5 \hspace{0.3em} {\rm AU}}\right)^{7/4}\exp\left(-\frac{t}{2.0 \hspace{0.3em} {\rm Myr}}\right) 
\Sigma_{\rm gas}({\rm 1.5 \hspace{0.3em}AU}, 0), \label{eq:sgas}
\end{equation}
where $\Sigma_{\rm gas}({\rm 1.5 \hspace{0.3em}AU}, 0)$ = 1,196 g cm$^{-2}$. 
The adopted decay time of the gas disk is consistent with observations (Fedele et al., 2010).
In Eq.~(\ref{eq:sgas}), the radial gradient of $\Sigma_{\rm gas}$ is chosen so that 
the rotation velocity of the gas is exactly the local Kepler velocity at the midplane. Thus, 
there is no radial drift due to gas drag if $e=i=0$. 
Aerodynamic gas drag and damping due to tidal interactions with the gas disk are taken into account
following the approach of Morishima et al. (2010), whereas Type I migration (Goldreich and Tremaine, 1980) is neglected. 
A model of planetesimal formation in turbulence (Chambers 2010) suggests a planetesimal 
mass even smaller than $m_0$ at 1.5 AU. To mimic smaller sizes of planetesimals in the gaseous disk, 
the drag force on the smallest planetesimal is enhanced by a factor of $(m_0/m_{\rm GI})^{2/3}$, 
where  $m_{\rm GI}$ is the actual planetesimal mass of interest,  and the enhancement factor 
smoothly decreases with increasing mass from $m_0$ to $10 m_0$  (see Morishima et al., 2010).
We set $m_{\rm GI}$ to be $10^{21}$ g,  which is close to that at 1.5 AU suggested by Chambers (2010).

For this study, the $N$-body code developed by Morishima et al. (2010) is used. In this code,
the mutual gravity of all bodies is calculated with a parallel-tree method,  
and the orbital integration is calculated with a mixed-variable symplectic integrator. 
Since this code can handle a large number of particles and take a large time step, 
it is applicable to any stage of planetary accretion.
The opening parameter for the tree method is 0.5 and the time step is 11.6 days. 
Each simulation was performed for a total duration of $\sim$ 14 Myr and took roughly two months using a single node with an eight-core processor on a supercomputer.

\subsection{Results}
Figures~1-3 show results for the simulation with nebular gas (Sim.~A).
Snapshots on the $a$-$e$ plane are shown in Fig.~1. 
The large embryos are displayed as red filled circles with horizontal branches with a length of $5 r_{\rm H}$ on each side.
We define an embryo as a body more massive than $10^{25}$ g ($\simeq m_{\rm tr}$ (Eq.~(\ref{eq:mtr})) for $b_{\rm e} = 10$), 
but bodies less massive than one-tenth of the most massive body's mass are excluded, as such a small body
behaves like a planetesimal under gravitational stirring by massive embryos.
As embryos grow with time, the number of planetesimals decreases and their eccentricities increase. 
At $t =$ 1 Myr, many orbital overlaps of embryos can be observed.  
Through collisions between embryos, two large embryos form. Their orbital separation 
is roughly 10$r_{\rm H}$ as found in Kokubo and Ida (1998, 2000).
At the end, their masses reach  80-85\% of the mass of Mars.

The cumulative number of bodies and the velocity distribution as a function of mass 
are shown in Fig.~2. While the power-law index for the mass distribution 
is initially $-2.5$,  it  gradually evolves to $\sim -2$ as a result of embryo growth.
The large embryos have low $e$ due to tidal interactions with gas and dynamical friction of surrounding planetesimals. 
For small planetesimals, $e$ decreases with decreasing $m$ due to gas drag. 
The analytic estimates of the velocity dispersion (Eq.~(\ref{eq:tile})) of the smallest bodies 
are shown on the left side as dotted lines and coincide well with those of the $N$-body simulation.

The masses of the two surviving embryos as a function of time are shown in the top panel of Fig.~3.
In the same panel, the impactor-to-embryo mass ratios ($\le 1$) are shown as diamonds for embryo-embryo collisions.
The total mass gained by embryo-embryo collisions, $f_{\rm e-e}$, relative to the mass of the embryo is shown in the middle panel
of Fig.~3 for each embryo. The number of embryos $N_{\rm e}$ and the mean separation normalized by the Hill radius $b_{\rm e}$
are shown in the bottom panel of Fig.~3.
Initially, $N_{\rm e}$ increases and $b_{\rm e}$ decreases with time. 
After $b_{\rm e}$ reduces to $\sim$ 5 around 1 Myr, embryo-embryo collisions start to occur frequently and 
$N_{\rm e}$ decreases while $b_{\rm e}$ increases. As a result, 
surviving embryos gain roughly half of their masses by embryo-embryo collisions.
These fractions decrease with time near the end of the simulation because all other 
embryos are swept up at 6 Myr and thereafter only small planetesimals collide with the embryos.
The fractions extrapolated to the end state where all planetesimals are completely swept up are 0.1-0.2.

On the top panel of Fig.~3, we also plot the theoretically expected mass evolution (Eq.~(\ref{eq:me})).
The velocity dispersion calculated from Eq.~(\ref{eq:tile}) increases with time and its upper limit is fixed to be the escape velocity of 
embryos in Eq.~(\ref{eq:me}). We adopt $b_{\rm e} = 9.1$ in Eq.~(\ref{eq:me}) so that $m_{\rm iso}$ = $M_{\rm d}/2$ = $10^{27}$ g;
this is the expected final mass of the two surviving embryos.
These conditions give $\tau_{\rm grow} = 3.06$ Myr in the beginning of the simulation.
The $N$-body simulation shows a good agreement with the analytic estimate up to $\sim$ 3 Myr.
Good agreement with the analytic model adopting a bimodal size distribution is reasonable as a large mass fraction is in smallest planetesimals in the simulation. 

However, after $\sim$ 3 Myr the simulation shows that the embryos grow by a factor of three slower than the analytic estimate.
There are several plausible reasons reducing the growth rate in the simulation. First, embryo-embryo collisions cease around 3 Myr 
while they are assumed to occur constantly in the analytical estimation (C is fixed to be 1.5 in Eq.~(\ref{eq:aa})).   
Cease of embryo-embryo collisions in the middle of planetary accretion is also seen in $N$-body simulations without the periodic boundary condition 
(Morishima et al., 2008).
Second,  the surface densities in the feeding zones of embryos are reduced relative to those near the boundaries of the annulus,
due to strong scattering by embryos.
A non-uniform spatial distribution of planetesimals is seen even without gas (Ida and Makino, 1993) 
but non-uniformity is enhanced with gas drag (Tanaka et al., 1997). 
This effect is already seen at $t=$ 2 Myr in Fig.~1 and also seen in simulations of Kokubo and Ida (2000). 
Third, the width of the feeding zone of an embryo increases with orbital eccentricities 
of planetesimals. Thus, during the late stage,  a non-negligible part of the feeding zone lies outside the simulated annulus.
This effect is likely to be relatively unimportant compared to the first and second effects.
Overall, the analytic model is likely to underestimate the timescale of sweeping up of remnant planetesimals during 
the late oligarchic growth stage.

Figures~4 and 5 show the same as Figs.~1 and 3 but for the simulation without nebular gas (Sim.~B).
Overall evolution of Sim.~B is similar to Sim.~A.
Due to a higher velocity dispersion in the absence of gas, however, 
embryo growth is significantly slower than in Sim.~A. 
The velocity dispersion is found to be roughly half of the escape velocity of the largest embryo throughout the simulation.
The final number of embryos is only one in Sim.~B  in contrast to two in Sim.~A. 
This is because orbits of embryos become more unstable 
without damping due to tidal interactions with the gas disk (Iwasaki et al., 2002; Kominami et al., 2002).
It is expected that  the embryo-embryo spacing becomes much larger without gas than with gas in the late stage
if the simulation annulus is much wider.
Related to this, the peak number of embryos for Sim.~B (Fig.~5) is only half of that for Sim.~A.
Despite these differences, the mass fraction of the largest embryo obtained by embryo-embryo collisions in Sim.~B is similar to those in Sim.~A.
The mass growth curve roughly coincides with the analytic estimation assuming that 
the velocity dispersion is half of the escape velocity of the largest embryo (Fig.~5) and $m_{\rm iso} = M_{\rm d} = 2 \times 10^{27}$ g ($b_{\rm e} = 14.4$ and $\tau_{\rm grow} = 15.63$ Myr).
The growth rate in the late stage of the simulation is likely to be lower than the analytic estimate for the same reasons mentioned for Sim.~A. 

\section{Evolution of Hf-W isotopes}
The Hf-W chronology provides strong constraints on core formation timing in planetary bodies 
(Jacobsen, 2005; Kleine et al., 2009).
$^{182}$Hf  decays to $^{182}$W with a half-life of 8.9 $\pm$ 0.1 Myr. 
Hf and W are both refractory elements and their relative abundances 
in bulk planetary bodies should be chondritic.
During core formation, lithophile Hf entirely remains in a silicate mantle
whereas siderophile W is preferentially partitioned into a metallic core.

The evolution of Hf-W isotopes is investigated using the results of the $N$-body simulations.  
Since the $N$-body simulations were halted before complete sweep up of planetesimals, 
we extrapolate the growth curves of the embryos (dashed curves in Figs.~3 and 5)
using Eq.~(\ref{eq:dsige}); the growth timescale $2/A$ of an embryo is derived using the masses of 
the embryo at $t =$ 10 Myr and at the end of the simulation ($t \sim$ 14 Myr) 
assuming $A$ is constant during this period. 
As an extreme comparison, we also calculate the evolution of the Hf-W isotopes
for a case without subsequent accretion after the end of the simulation
assuming that remaining planetesimals are suddenly dispersed.
 
\subsection{Formulation}
We make the following assumptions:
\begin{enumerate}
\item The abundances of Hf and W in all bulk bodies are chondritic.
\item The metal-silicate partition coefficient for W, $D^{\rm W}$, is independent of time and space. 
The silicate fraction $y$ and the complementary metal fraction $1-y$ are also constant for all bodies.
\item All bodies experience core formation at the beginning of the solar system due to $^{26}$Al radiogenic heating and their cores and mantles are fully equilibrated at this time
(the timing of the core-mantle differentiation is found to have little effect).
\item During a collision,  the core of the target (the more massive body) is not involved in metal-silicate equilibration. The target's mantle and the impactor's core are 
partially involved in equilibration and the fractions of involved masses are defined to be $k_{\rm tm}$ and $k_{\rm ic}$, respectively.
\item After equilibration, the impactor's core merges with the target's core.
The target's mantle is well mixed and homogenized immediately after the core merging,  as well as the core,
but there is no material exchange between the core and the mantle.  
\end{enumerate}

The W isotope ratio $^{182}$W/$^{183}$W for a sample relative to that for the CHondritic Uniform Reservoir (CHUR) is defined as  
\begin{equation}
\Delta \epsilon (t) = \left( \frac{(^{182}W/^{183}W)(t)}{(^{182}W/^{183}W)_{\rm CHUR}(t)} - 1\right) \times 10^4,
\end{equation}
where $t$ is the time measured from  the beginning of the solar system (4,568 Myr ago; Kleine et al., 2009). 
The value for the present-day Mars is estimated from Shergottites to be $\Delta \epsilon = 2.68 \pm 0.26$ (Dauphas and Pourmand, 2011).

The radiogenic change of $\Delta \epsilon$ during a time interval between two collisions (the first and the second collisions occurring at $t1$ and $t2$) 
is given by (Jacobsen 2005)
\begin{equation}
\Delta \epsilon(t2) -  \Delta \epsilon(t1) =   C_{\rm W} \left(e^{-\lambda t1} - e^{-\lambda t2} \right), \label{eq:ee1}
\end{equation}
where  $\lambda$ is the decay constant of ${}^{182}$Hf.
The coefficient $C_{\rm W}$ is given as
\begin{equation}
C_{\rm W} = q_{\rm W} \left(\frac{\rm {}^{182}Hf}{\rm {}^{180}Hf}\right)_{\rm CHUR}^{t=0}f^{\rm Hf/W}, \label{eq:cw}
\end{equation}
where
\begin{equation}
q_{\rm W} =  10^4\left( \frac{\rm {}^{180}Hf }{\rm {}^{182}W}\right)_{\rm CHUR}^{t=0},
\end{equation}
and
\begin{equation}
f^{\rm Hf/W} =  \left( \frac{\rm ({}^{180}Hf/{}^{183}W)}{\rm ({}^{180}Hf/{}^{183}W)_{\rm CHUR}} - 1\right).
\end{equation}
Strictly speaking, $q_{\rm W}$ should be the value at time $t$ but can be well approximated as a constant for the Hf-W system.
The values from Dauphas and Pourmand (2011; and references therein) are used:
$q_{\rm W} = 1.07 \times 10^4$, 
$({\rm {}^{182}Hf}/{\rm {}^{180}Hf})_{\rm CHUR}^{t=0} = 9.72 \times 10^{-5}$,
$f^{\rm Hf/W} = 3.38 \pm 0.56$, and $\lambda = 0.0779$ Myr$^{-1}$.
Only the uncertainty in $f^{\rm Hf/W}$ is taken into account, as it has the largest effect on the estimated accretion timescale.
With assumptions 2 and 3, $f^{\rm Hf/W}$ is time-independent for all bodies. Additionally using assumption 1,  $f^{\rm Hf/W}$ is given as (Jacobsen, 2005)
\begin{equation}
f^{\rm Hf/W} = \frac{(1-y)D^{\rm W}}{y}. \label{eq:fhfw}
\end{equation}

A change in $\Delta \epsilon$ due to a collision, ignoring radiogenic decay during the collision, is 
\begin{equation}
\Delta \epsilon_{1} =  f_{\rm t} \Delta \epsilon_{0} + f_{\rm i} \Delta \epsilon_{\rm i}, \label{eq:ee2}
\end{equation}
where $\Delta \epsilon_{0}$ and $\Delta \epsilon_{1}$ are  $\Delta \epsilon$ of the target before and after 
the collision and $\Delta \epsilon_{\rm i}$ is $\Delta \epsilon$ of the impactor.
The reduction factor $f_{\rm t}$ represents how much ${}^{182}$W in the target's mantle is transported to the target's core
by the impactor's core while $f_{\rm i}$ represents how much the super chondritic impactor's mantle is added to the target's mantle.
These factors are given as (see Appendix)
\begin{equation}
f_{\rm t} = \frac{(k_{\rm tm} + g) + (1-k_{\rm tm})k_{\rm ic}f^{\rm Hf/W}g}{(k_{\rm tm} + g + k_{\rm ic}f^{\rm Hf/W}g)(1+g)} \label{eq:ft}
\end{equation} 
and 
\begin{equation}
f_{\rm i} = \frac{(1-k_{\rm ic})(k_{\rm tm} + g)g}{(k_{\rm tm} + g + k_{\rm ic}f^{\rm Hf/W}g)(1+g)},\label{eq:fi}
\end{equation} 
where $g (\le 1)$ is the mass ratio of the impactor to the target. 
In the case of $k_{\rm ic} = 1$, $f_i $ will become zero. Thus, a change in  $\Delta \epsilon$ does not depend on the isotopic fractionation history of the impactor. 
For $k_{\rm ic} = k_{\rm tm}  = 1$, Eq.~(\ref{eq:ft}) is reduced to Eq.~(A6) of Kleine et al. (2009). 
This case is called the mantle equilibration scenario in Nimmo and Agnor (2006).
The core-merging scenario of Nimmo and Agnor (2006) corresponds to a case with $k_{\rm ic} = 0$ and $k_{\rm tm}  = 1$.

Time evolution of $\Delta \epsilon$ of an embryo is calculated using Eqs.~(\ref{eq:ee1}) and (\ref{eq:ee2}).
After the end of an $N$-body simulation,  
impacts to the embryo are assumed to occur every $\Delta t$ and 
the impactor mass is given by $m_{\rm e}(t) - m_{\rm e}(t-\Delta t)$. We adopt a somewhat large $\Delta t (= 10^6$ yr) 
to represent remaining medium-sized impactors ($g \sim 0.01$).
For $k_{\rm ic} < 1$,  $\Delta \epsilon$'s of all bodies are calculated while 
we assume that $\Delta \epsilon_{\rm i} = \Delta \epsilon_0$ after the end of the simulation.

\subsection{Results}
We first show the isotopic evolution for the case of perfect equilibration ($k_{\rm tm} = k_{\rm ic} = 1$). 
Next,  the effect of partial equilibration of the target's mantle is examined ($k_{\rm tm} \le 1$). 
Third, the effect of partial equilibration of the impactor's core is studied ($k_{\rm ic} \le 1$).
Finally,  partial equilibration of both the target's mantle and the impactor's core 
is taken into account. 

\subsubsection{Perfect equilibration}
Figure~6 shows the time evolution of $\Delta \epsilon$ for the case of perfect equilibration ($k_{\rm tm} = k_{\rm ic} = 1$).
Results for two embryos from simulation A (red and blue solid curves) and for one embryo 
from simulation B (black curve) are shown. The dashed curves show a case without accretion after the end of the $N$-body simulation.
With increasing time, $\Delta \epsilon$ increases due to radiogenic decay of $^{182}$Hf to $^{182}$W, 
while impacts reduce $\Delta \epsilon$ as $^{182}$W is transported from the 
mantle to the core. The larger the impactor-to-target ratio $g$, the larger the decrease in $\Delta \epsilon$.
After $\sim$ 20 Myr, $\Delta \epsilon$ starts to decrease because decrease due to impacts 
exceeds increase due to radiogenic decay.
The present-day values of $\Delta \epsilon$ become less than 0.3 for all cases 
and are much lower than the martian value.  
If accretion is halted at the end of the $N$-body simulations, 
the present-day values of $\Delta \epsilon$ become much larger but still lower than the martian value.

\subsubsection{Partial equilibration of the target's mantle}
If the impactor size is much smaller than the target size,  the entire target's mantle may not be involved in
the metal-silicate equilibration. This effect is modeled by assuming that 
the volume of a portion of the target's mantle interacting with the impactor's core is proportional to the impactor's volume: 
\begin{equation}
k_{\rm tm} = {\rm min}[c_{\rm tm} g, 1.0]. \label{eq:ktm}
\end{equation}
where $c_{\rm tm}$ is the proportionality coefficient (ignoring a factor of $y$).  A similar form is also used in Sasaki and Abe (2007).  
If the target has a liquid magma ocean on the surface, the volume of the interacting portion   
is probably given by the product of the crater area and the magma ocean depth.
In this case, $c_{\rm tm}$ is likely to be much larger than unity particularly for small impactors.
On the other hand, if the target surface temperature is lower than the solidus temperature of the rocks composing the mantle,
the impact-induced melt in the target's mantle may be taken as the interacting portion.
The volume of the impact-induced melt primarily depends on impact velocity,
and if the impact velocity is about the escape velocity of Mars,  only the isobaric core, whose volume is a few times of the impactor volume ($c_{\rm tm} \sim 3$),
results in melting (Tonks and Melosh, 1993, 1992). 

Figure~7 shows the present-day value of $\Delta \epsilon$ as a function of $c_{\rm tm}$.
The effect of partial equilibration is found to be unimportant for large $c_{\rm tm}$.
Inserting $k_{\rm tm} = c_{\rm tm} g$ into Eq.~(\ref{eq:ft}) and adopting $c_{\rm tm} \gg f^{\rm Hf/W}$, we have $f_{\rm t} \simeq (1-f^{\rm Hf/W}g)/(1+g)$.
This is equivalent to the case of perfect equilibration if $g \ll 1$. 
This indicates that the tungsten concentration in the impactor's core tends to saturate if the volume of the equilibrating silicate portion exceeds the criterion, 
$c_{\rm tm} \sim f^{\rm Hf/W}$. 
The present-day value of $\Delta \epsilon$ becomes much larger if $c_{\rm tm} < f^{\rm Hf/W}$ ($f_{\rm t} \simeq (1-c_{\rm tm}g)/(1+g)$ for $c_{\rm tm} \ll f^{\rm Hf/W}$).
Such a case is unlikely for Mars because its $f^{\rm Hf/W}$ is low and $c_{\rm tm}$ is likely to be at least $\sim 3$ as discussed above.
The effect of partial equilibration may be much more important for the Earth which has a much larger $f^{\rm Hf/W}$; 
the Hf/W ratio of 25.8 (K\"{o}nig et al., 2011) gives $f^{\rm Hf/W}$ = 25.3 using the Hf/W ratio of 0.98 for CI chondrites.

\subsubsection{Partial equilibration of the impactor's core}
Cases of a constant $k_{\rm ic}$ ($\le 1$) for all impacts are first considered while keeping $k_{\rm tm} = 1$.
The reductions of  $\Delta \epsilon$ after impacts are much smaller than in the case of perfect equilibration.
Eq.~(\ref{eq:ft}) shows that $f_{\rm t} = (1+g+k_{\rm ic}f^{\rm Hf/W}g)^{-1}$. 
Thus, the effect of partial equilibration of the impactor's core is equivalent to reducing $f^{\rm Hf/W}$ to $k_{\rm ic}f^{\rm Hf/W}$, 
or to reducing the metal-silicate partition coefficient $D^{\rm W}$ to $k_{\rm ic}D^{\rm W}$.
In addition, the disequilibrated impactor's mantle is added to the target's mantle (Eq.~(\ref{eq:fi})).
This leads to an additional increase in $\Delta \epsilon$. 

Figure~8 shows the present-day value of $\Delta \epsilon$ as a function of $k_{\rm ic}$.
As discussed, $\Delta \epsilon$ increases with decreasing $k_{\rm ic}$. 
In the limit of $k_{\rm ic} \rightarrow 0$, $\Delta \epsilon = C_{\rm W}$ (Eq.~(\ref{eq:cw})) regardless
of accretion history.  
It is found that $\Delta \epsilon$ for present-day Mars is fulfilled with 
$k_{\rm ic}$ less than 0.1. This value is even smaller than the lower limit (0.36) suggested for the Earth (Rudge et al., 2010). 
If accretion is halted at the end of the $N$-body simulation with gas, 
$k_{\rm ic}$ needs to range from 0.2 to 0.7.

\subsubsection{A model with a deep magma ocean}
In this section, we consider a case where Mars has a deep magma ocean during its accretion. Whether this assumption is reasonable 
is discussed in the next section.
As found in hydrodynamic modeling (Deguen et al., 2011; Samuel, 2012),
$k_{\rm ic}$ is likely to decrease with increasing impactor size.
As a nominal model, we adopt the following form:
\begin{equation}
k_{\rm ic} = {\rm min}[c_{\rm ic} g^{-1}, 1], \label{eq:kic}
\end{equation}
where $c_{\rm ic}$ is the proportional coefficient and we adopt $c_{\rm ic}  = 0.01$. 
This means that large impactors only partially equilibrate with the target  whereas
smaller impactors ($g \le 0.01$) fully equilibrate. This model is probably close to the condition 
derived by Samuel (2012), who found that impactors with sizes smaller than the depth of 
the target's magma ocean fully equilibrate. 
We also adopt partial equilibration of the target's mantle
using Eq.~(\ref{eq:ktm}) with $c_{\rm tm} = 10$.  As discussed in Sec.~4.2.2, this is likely to cause 
only a very small difference from the case of $k_{\rm tm} = 1$.
 
The evolution curves of $\Delta \epsilon$ are shown in Fig.~9.  
There is no large decrease in $\Delta \epsilon$ at impacts unlike in Fig.~6 because 
$k_{\rm ic}$ for large impactors is very small. 
Nevertheless, the increase in $\Delta \epsilon$ as compared with those for 
the cases of perfect equilibration is rather small,
because more than half of an embryo mass is delivered by small impactors 
and their cores fully equilibrate with the embryo mantle.
The present-day values of $\Delta \epsilon$ are close to the values obtained with $k_{\rm ic} = 0.5-0.7$ 
in Fig.~8. This is reasonable as the fraction of small impactors is roughly two-thirds
as shown in Figs.~3 and 5.
In any cases, the present-day values of $\Delta \epsilon$ are much lower than the 
martian value.
If accretion is halted at the end of the $N$-body simulation with gas,
the present-day values of $\Delta \epsilon$ becomes barely consistent with 
the martian value.
  
So far, we have fixed the value of  $f^{\rm Hf/W}$ to be 3.38. 
Nimmo and Kleine (2007) showed that 
different values of $f^{\rm Hf/W}$ result in very different accretion timescales.  
Thus,  the uncertainty of $f^{\rm Hf/W}$ needs to be taken into account, although
it is now much smaller owing to the work of Dauphas and Pourmand (2011).
Figure~10 shows $\Delta \epsilon$ as a function of $f^{\rm Hf/W}$.
It is found that $\Delta \epsilon$ is almost independent of $f^{\rm Hf/W}$ (solid lines).
Increase of $\Delta \epsilon$ with time due to radiogenic decay
is simply proportional to $f^{\rm Hf/W}$ (Eq.~(\ref{eq:ee1})).
On the other hand, the decrease of $\Delta \epsilon$ due to an impact
also increases with increasing $f^{\rm Hf/W}$ (see Eq.~(\ref{eq:ft}) and Sec.~4.2.2).
If the accretion timescale is longer than the radiogenic decay timescale, both 
effects roughly compensate each other and the present-day $\Delta \epsilon$ depends 
very weakly on $f^{\rm Hf/W}$.  On the other hand, if accretion rapidly completes as in the cases shown by dashed lines in Fig.~10,
the present-day $\Delta \epsilon$ is primarily determined by radiogenic decay
so $\Delta \epsilon$ is roughly proportional to $f^{\rm Hf/W}$.  Similar dependence and independence 
can be found if  $f^{\rm Hf/W}$ is varied in calculations shown in Fig.~6.

\section{Discussion}

In Sec.~4.2.4, we showed that 
as long as a Mars-analog has a deep magma ocean during its accretion and 
the final stage of accretion is prolonged, 
the present-day $\Delta \epsilon$ turns out to be much lower than the martian value. 
This may indicate that accretion of Mars proceeded and terminated much more rapidly than in our model calculations, because
(1) the solid surface density was larger, (2) the gas surface density was larger, or 
(3) the planetesimals size was smaller than what we have assumed. 
 For the first possibility, the terrestrial planets might have accreted in a narrow annulus around 1 AU,
 and Mars is ejected to the current location (Hansen, 2009; Morishima et al., 2008; Walsh et al., 2011). 
 Although this scenario is favorable for the strong radial mass concentration of the terrestrial planets, 
 the Earth is likely to accrete too rapidly to explain its $\Delta \epsilon$ and it may be difficult to produce 
 the large difference in the oxidization states 
 (represented by $f^{\rm Hf/W}$ and the FeO content) between Earth and Mars (but see Morbidelli and Rubie (2012)). 
 For the second possibility, the high gas density may be achieved in a low turbulent viscosity region called 
 the dead zone (Gammie, 1996; Morishima, 2012).  
 For the third possibility,  growth of embryos is significantly accelerated  due to accretion of 
 fragments of planetesimals (Chambers, 2006; Kobayashi et al., 2010; Wetherill and Stewart, 1993) 
 or small pebbles that are precursors of planetesimals (Lambrechts and Johansen, 2012; Morbidelli and Nesvorny, 2012).
A quantitative discussion assuming perfect metal-silicate equilibration can be found in Kobayashi and Dauphas (2012).
The mechanisms listed above also help to terminate accretion quickly because
Mars is ejected to a low surface density region, or remnant planetesimals are removed by strong gas drag. Perturbations from 
giant planets also help to disperse remnant planetesimals (e.g., Morishima et al., 2010).  
Unfortunately, these effects cannot be properly evaluated in local simulations with the periodic boundary condition, although
it is evident that rapid dispersal of remnant planetesimals significantly enhances tungsten anomaly (dashed lines in Fig.~6 and 9).

Alternative to the rapid accretion discussed above, 
the degree of metal-silicate equilibration might have been very low ($k_{\rm ic} \le 0.1$, see Sec.~4.2.3) while 
accretion of Mars was as slow as in our simulations (see also Mezger et al., 2012).
Such a low degree of equilibration might be feasible if 
accretion still proceeds while the global magma ocean starts to solidify.
However, geochemical evidence seems to suggest that Mars gained most of its mass while it had a deep magma ocean.
The most recent study of partitioning of the siderophile elements indicates a deep ($>$ 1,000 km) global magma ocean on Mars (Righter and Chabot, 2011),
contrary to a shallow magma ocean suggested by previous studies (Righter and Drake, 1996; Kong et al., 1999).
A deep magma ocean of Mars is also indicated from a differentiation model (Elkins-Tanton et al., 2005) that can produce magma source regions consistent with 
element data for SNC meteorites and the martian crust.
For Mars-size bodies, both impact heating and radiogenic heating due to the decay of $^{26}$Al
are likely to contribute to formation of magma oceans (Rubie et al., 2007; \v{S}r\'{a}mek et al., 2012). 
Numerical models suggest that the presence of a magma ocean on Mars is very likely (Monteux et al., 2010; Ricard et al., 2009; \v{S}r\'{a}mek et al., 2012)
but its depth and duration depends on various uncertain parameters. An important parameter is impactor sizes which can be directly obtained from $N$-body simulations. 
Thermal evolution of the embryos in our $N$-body simulations is investigated in a companion paper (Golabek et al., in preparation). 
  
  
A late stage giant impact is favorable for the origin of the geological dichotomy of Mars 
(Golabek et al., 2011; Marinova et al., 2008; Nimmo et al., 2008; Wilhelms and Squyres, 1984).
The impact is unlikely to have significantly reset the chronometer probably because
the impactor size was not large enough to stir the entire martian mantle 
or emulsification of the impactor's core was inefficient. 
  
  
\section{Summary}
We have performed  high-resolution $N$-body simulations for oligarchic growth of Mars,
using the body supply boundary condition which keeps the total solid mass  in the simulation annulus constant.  
One simulation in the gas-free environment and another simulation with nebular gas were carried out.
The surface densities of solid and gas are larger than those for the minimum mass solar nebula model by only 30$\%$.
The time evolution of the embryo mass in nebular gas coincides well with that of 
the analytic model of Chambers (2006) until the middle of accretion.
However, the $N$-body simulations show much slower accretion than the analytic model in the late stage. 
The mass delivered by embryo-embryo collisions relative to the embryo mass is about half at maximum in the middle of accretion. 
This fraction decreases to 0.1-0.2 when planetesimals are completely swept up because 
only small planetesimals collide with the embryos during the late stage of accretion.

We have calculated the Hf-W isotopic evolution of the embryos, using accretion histories from the $N$-body simulations. 
It is likely that only a part of the target's mantle is involved in the equilibration 
if the impactor size is small while only a small fraction of the impactor's core 
is involved for a large impactor. 
These effects are modeled and examined using the output of the $N$-body simulations.
It was found that as long as the target has a deep liquid magma ocean during its accretion, the effects of imperfect equilibration are small. 
The effect of partial involvement of the target's mantle is small as long as 
 the volume of the target's mantle portion involved in equilibration relative to the impactor volume
is larger than $f^{\rm Hf/W}$. This is likely to be the case for Mars as its $f^{\rm Hf/W}$ value is low. 
The effect of partial involvement of the impactor's core is also small because a large fraction of the embryo mass is delivered by small impactors,
which are likely to fully equilibrate in the deep magma ocean of the embryo (Samuel, 2012). 

We found that
with a prolonged accretion suggested from our simulations ($\sim$ 100 Myr), 
the high value of the martian tungsten anomaly can be achieved only if 
less than 10\% of each impactor's core re-equilibrates with the martian mantle. 
This indicates that  growth of Mars proceeded rapidly due to solid and gas surface 
densities significantly larger than those for the minimum mass solar nebula or due to accretion of small fragments or pebbles.

\section*{Acknowledgements}
We thank an anonymous reviewer and Alessandro Morbidelli for their constructive comments and 
Hiroshi Kobayashi for fruitful discussions and sending his preprint.
This research was partly carried out at the Jet Propulsion Laboratory, California Institute of Technology, 
under contract with NASA. Government sponsorship is acknowledged. G.J.G. was supported by SNF grant PBEZP2-134461.
H.S. acknowledges the funds from the Stifterverband f\"{u}r Deutsche Wissenschaft. 
The $N$-body simulations were carried out on supercomputer Schr\"{o}dinger at University of Zurich. 

\section*{Appendix: changes of tungsten isotope concentrations during a collision}
The mass concentrations of the radiogenic and stable isotopes of tungsten are
defined as  $^{182}$W and $^{183}$W. 
Let us assume that a change in $^{182}$W (or $^{183}$W) due to a collision is given as: 
\begin{equation}
{}^{182}{\rm W}_{\rm tm, 1}  =  f_{\rm t}  {}^{182}{\rm W}_{\rm tm, 0} +  f_{\rm i} {}^{182}{\rm W}_{\rm im}  + f_{\rm c} {}^{182}{\rm W}_{\rm CHUR},
\end{equation}
where  $f_{\rm t}$, $f_{\rm i}$, and $f_{\rm c}$ are the coefficients derived below, 
the subscripts tm, im, and CHUR represent the target's mantle, the impactor's mantle, and the CHondritic Uniform Reservoir,
and the subscript numbers 0 and 1 for the target's mantle mean before and after the collision, respectively.
With assumptions 2 and 3 from Sec.~4.1,  the concentration of the stable isotope $^{183}$W does not change due to a collision even for $k_{\rm ic} < 1$ or  $k_{\rm tm}  < 1$ (proved below).
This leads to 
\begin{equation}
{}^{183}{\rm W}_{\rm tm, 1}  =  {}^{183}{\rm W}_{\rm tm, 0}  =  {}^{183}{\rm W}_{\rm im}  = \frac{f_{\rm c}}{1-f_{\rm t} -f_{\rm i}}  {}^{183}{\rm W}_{\rm CHUR}, \label{eq:w183}
\end{equation}
and 
\begin{equation}
\Delta \epsilon_{1} =  f_{\rm t} \Delta \epsilon_{0} + f_{\rm i} \Delta \epsilon_{\rm i},
\end{equation}
where $\Delta \epsilon_{1}$ and $\Delta \epsilon_{0}$ are $\Delta \epsilon$ of the target's mantle after and before the collision 
and $\Delta \epsilon_{\rm i}$ is for the impactor's mantle.

In the following, the factors $f_{\rm t}$, $f_{\rm i}$, and $f_{\rm c}$ are derived.  
The total mass involved in equilibration is 
\begin{equation}
M_{\rm eq} = k_{\rm tm}y_{\rm t} M_{\rm t} + y_{\rm i} M_{\rm i} + k_{\rm ic} (1-y_{\rm i}) M_{\rm i},
\end{equation}
where $M_{\rm t}$ and $M_{\rm i}$ are the masses of the target and the impactor, and 
$y_{\rm t}$ and $y_{\rm i}$ are the silicate fractions of the target and the impactor.
Since the bulk composition of the impactor is assumed to be chondritic, we have
\begin{equation}
{}^{182}{\rm W}_{\rm CHUR} = y_{\rm i} {}^{182}{\rm W}_{\rm im} + (1-y_{\rm i}) {}^{182}{\rm W}_{\rm ic}, \label{eq:imc}
\end{equation}
where the subscript ic denotes the impactor's core. 
Using Eq.~(\ref{eq:imc}), 
the total mass  of equilibrated ${}^{182}{\rm W}$ supplied by the impactor relative to the impactor mass is given as
\begin{eqnarray}
{}^{182}{\rm W}_{\rm i,eq} &=& y_{\rm i}{}^{182}{\rm W}_{\rm im} + k_{\rm ic}(1-y_{\rm i}) {}^{182}{\rm W}_{\rm ic} \nonumber \\
&=& (1- k_{\rm ic})y_{\rm i} {}^{182}{\rm W}_{\rm im} + k_{\rm ic}{}^{182}{\rm W}_{\rm CHUR}.
\end{eqnarray}
The concentration ${}^{182}{\rm W}_{\rm eq, 0} $ averaged over the masses involved in the equilibration is given by
\begin{equation}
 {}^{182}{\rm W}_{\rm eq, 0} = \frac{1}{M_{\rm eq}}(k_{\rm tm}y_{\rm t}M_{t}{}^{182}{\rm W}_{\rm tm,0}  + M_{\rm i}{}^{182}{\rm W}_{\rm i,eq}) \label{eq:m0}
 \end{equation}
 The silicate fraction of the equilibrated masses is 
 \begin{equation}
 y_{\rm eq} = \frac{k_{\rm tm}y_{\rm t}M_{\rm t} + y_{\rm i}M_{\rm i}}{M_{\rm eq}} \label{eq:ym}
  \end{equation} 
 After the equilibration, the concentration ${}^{182}{\rm W}_{\rm eq, 1}$ of the equilibrated silicate portion is (Nimmo and Agnor, 2006) 
  \begin{equation}
 {}^{182}{\rm W}_{\rm eq, 1} =  \frac{{}^{182}{\rm W}_{\rm eq, 0}}{y_{\rm eq}+(1-y_{\rm eq})D^{\rm W}} \label{eq:m1}
   \end{equation} 
The fraction of ${}^{182}{\rm W}_{\rm tm, 1}$ for the entire target's mantle is given by averaging the equilibrated  and
 non-equilibrated portions as 
\begin{equation}
 {}^{182}{\rm W}_{\rm tm,1} = \frac{k_{\rm tm}  y_{\rm t} M_{\rm t} +  y_{\rm i} M_{\rm i}}{ y_{\rm t} M_{\rm t} +  y_{\rm i}M_{\rm i}} {}^{182}{\rm W}_{\rm eq, 1} 
 +  \frac{(1-k_{\rm tm})  y_{\rm t} M_{\rm t}}{ y_{\rm t}M_{\rm t} +  y_{\rm i}M_{\rm i}} {}^{182}{\rm W}_{\rm tm, 0}
 \label{eq:w1} 
 \end{equation} 
 From Eqs.~(\ref{eq:m0})-(\ref{eq:w1}), we have
\begin{equation}
f_{\rm t} = \frac{C_1 + (1-k_{\rm tm})C_2}{(C_1 + C_2)C_3}y_{\rm t}M_{\rm t}, \label{eq:fta}
\end{equation} 
\begin{equation}
f_{\rm i} = \frac{(1-k_{\rm ic})C_1}{(C_1 + C_2)C_3}y_{\rm i}M_{\rm i},\label{eq:fia}
\end{equation}
\begin{equation}
f_{\rm c} = \frac{k_{\rm ic}C_1}{(C_1 + C_2)C_3}M_{\rm i},\label{eq:fca}
\end{equation} where
\begin{equation}
C_1= k_{\rm tm}y_{\rm t}M_{\rm t}+ y_{\rm i}M_{\rm i},
\end{equation} 
\begin{equation}
C_2= k_{\rm ic}D^{\rm W}(1-y_{\rm i})M_{\rm i},
\end{equation} 
\begin{equation}
C_3= y_{\rm t}M_{\rm t}+ y_{\rm i}M_{\rm i}.
\end{equation}
For $y = y_{\rm i} = y_{\rm t}$, Eqs.~(\ref{eq:fta}) and (\ref{eq:fia}) are reduced to Eqs.~(\ref{eq:ft}) and (\ref{eq:fi}), respectively, using Eq.~(\ref{eq:fhfw}).

Eq.~(\ref{eq:w1}) is also applicable to the stable isotope  ${}^{183}{\rm W}$. Consider the case where both the impactor and the target body have not experienced any collision but have experienced core formation with full metal-silicate equilibration.
For these bodies with $y = y_{\rm i} = y_{\rm t}$, we have 
\begin{equation}
{}^{183}{\rm W}_{\rm tm, 0}  =  {}^{183}{\rm W}_{\rm im}  = \frac{{}^{183}{\rm W}_{\rm CHUR}}{y + (1-y)D^{\rm W}}.
\end{equation}  
Inserting this equation into  Eq.~(\ref{eq:w1}), we obtain ${}^{183}{\rm W}_{\rm tm, 1} =  {}^{183}{\rm W}_{\rm tm, 0}$ (Eq.~(\ref{eq:w183})).
This relationship holds for subsequent collisions.
Therefore, the concentration of the stable isotope is not changed by collisions, as long as $D^{\rm W}$ and $y$ are constant.
     
\section*{REFERENCES}
\begin{description}
\item
Chambers, J., 2006.
A semi-analytic model for oligarchic growth.
Icarus 180, 496--513.

\item 
Chambers, J., 2010.
Planetesimal formation by turbulent concentration.
Icarus 208, 505--517.

\item
Dahl, T.W., Stevenson, D.J., 2010. 
Turbulent mixing of metal and silicate during planet accretion and interpretation of the Hf-W chronometer. 
Earth Planet. Sci. Lett. 295, 177--186.

\item 
Dauphas, N.,  Pourmand, A., 2011.
Hf-W-Th evidence for rapid growth of Mars and its status as a planetary embryo.
Nature 473, 489--492.

\item
Deguen, R., Olson, P., Cardin, P., 2011. 
Experiments on turbulent metal-silicate mixing in a magma ocean. 
Earth Planet. Sci. Lett. 310, 303--313.
     
\item 
Elkins-Tanton, L.T., Hess, P.C., Parmentier,  E.M., 2005.
Possible formation of ancient crust on Mars through magma ocean processes.   
J. Geophys. Res. 110, E12S01.
     
\item
Fedele, D., van den Ancker, M.E., Henning, Th., Jayawardhana, R., Oliveira, J.M., 2010.     
Timescale of mass accretion in pre-main-sequence stars.
Astron. Astrophys. 510, A72.     
     
     
\item 
Gammie, C.F., 1996.
Layered accretion in T Tauri disks.
Astrophys. J., 457, 355--362.     
     
\item
Golabek, G.J., Keller, T., Gerya, T.V., Zhu, G., Tackley, P.J., Connolly, J.A.D., 2011.
Origin of the martian dichotomy and Tharsis from a giant impact causing massive magmatism.
Icarus 215, 346--357.     
     
\item
Goldreich, P., Ward, W.R. 1973.
The formation of planetesimals.
Astrophys. J. 183, 1051-1062.

\item Goldreich, P., Tremaine, S. 1980.
Disk-satellite interactions.
Astrophys. J. 241, 425--441.              
     
 \item
Hayashi, C. 1981.
Structure of the solar nebula, growth and decay of magnetic fields and effects of magnetic and turbulent viscosities on the nebula.
Suppl. Prog. Theoret. Phys. 70,  35--53.
    
\item
Hansen, B., 2009.
Formation of the terrestrial planets from a narrow annulus.
Astrophys. J. 703, 1131--1140.    
    
\item     
Ichikawa, H., Labrosse, S., Kurita, K., 2010.
Direct numerical simulations of an iron rain in the magma ocean.
J. Geophys. Res. 115, B01404.

\item 
Ida, S., Makino, J., 1993.
Scattering of planetesimals by a protoplanet: slowing down of runaway growth.    
Icarus  106, 210--227. 
    
\item
Iwasaki, K., Emori, H., Nakazawa, K., Tanaka,H., 2002.
Orbital stability of a protoplanet system under a drag force proportional to the random velocity.
Publ.  Astron. Soc. Japan 54, 471--479.   
    
\item
Jacobsen, S.B.,  2005.
The Hf-W isotopic system and the origin of the Earth and Moon. 
Ann. Rev. Earth Planet. Sci. 33, 531--570.    
    
\item 
Kendall, J.D., Melosh, H.J., 2012.
Fate of iron cores during planetesimal impacts.
Lunar Planet. Inst. Sci. Conf. Abstr. 41, 2699.    
    
\item 
Kleine, T., et al.,  2009. 
Hf-W chronology of the accretion and early evolution of asteroids and terrestrial planets.
Geochim. Cosmochim. Acta 73,  5150--5188.    
    
\item 
Kobayashi, H., Tanaka, H., Krivov, A.V., Indaba, S., 2010.
Planetary growth with collisional fragmentation and gas drag.
Icarus 209, 836--847.   
    
\item
Kobayashi, H., Dauphas, N., 2012.
Small planetesimals formed Mars.
Submitted to Icarus.
        
\item
Kokubo, E.,  Ida, S. 1998.
Oligarchic growth of protoplanets.
Icarus 131, 171--178.

\item 
Kokubo, E., Ida, S. 2000.  
Formation of protoplanets from planetesimals in the Solar nebula.
Icarus 143, 15--27.

\item
Kominami, J., Ida, S. 2002.
The effect of tidal interaction with a gas disk on formation of terrestrial planets.
Icarus 157, 43--56.

\item
Kong, P., Ebihara, M., Palme, H., 1999.
Siderophile elements in Martian meteorites and implications for core formation in Mars. 
Geochim. Cosmochim. Acta 63, 1865--1875. 

\item
K\"{o}nig, S., M\"{u}nker, C., Hohl, S., Paulick, H., Barth, A.R., Lagos, M., Pf\"{a}nder, J., B\"{u}chl, A., 2011.
The Earth's tungsten budget during mantle melting and crust formation.
Geochim. Cosmochim. Acta 75, 2119--2136.
 
\item
Lambrechts, M., Johansen, A., 2012.
Rapid growth of gas-giant cores by pebble accretion.
Astron. Astrophys. 544, A32. 
 
\item
Lissauer, J.J., 1987. 
Timescales for planetary accretion and the structure of the protoplanetary disk. 
Icarus 69, 249--265.

\item 
Marinova, M.M., Aharonson, O., Asphau, E., 2008.
Mega-impact formation of the Mars hemispheric dichotomy.
Nature 453, 1216--1219.

\item 
Mezger, K., Debaille, V., Kleine, T., 2012.
Core formation and mantle differentiation on Mars.
Space Sci. Rev. 174, 27--48.

\item
Monteux, J., Ricard, Y., Coltice, N., Dubuffet, F., Ulvrov\'{a}, M., 2009. 
A model of metal-silicate separation on growing planets. 
Earth Planet. Sci. Lett. 287, 353--362.

\item 
Morbidelli, A., Nesvorny, D., 2012.
Dynamics of pebbles in the vicinity of a growing planetary embryo: hydro-dynamical simulations.
Astron. Astrophys. 546, A18.

\item 
Morbidelli, A., Rubie, D.C., 2012.
Dynamical and chemical modeling of terrestrial planet accretion.
Goldschmidt 2012 conf. session 01e1 abst.

\item 
Morishima, R., Schmidt, M.W., Stadel, J., Moore, B. 2008.
Formation and accretion history of terrestrial planets from runaway growth through to late time: Implications for orbital eccentricity.
Astrophys. J. 685, 1247--1261.

\item 
Morishima, R., Stadel, J., Moore, B. 2010.
From planetesimals to terrestrial planets: $N$-body simulations including the effects of nebular gas and giant planets.
Icarus 207, 517--535.

\item 
Morishima, R., 2012.
Gap opening beyond dead zones by photoevaporation.
Mon. Not. R. Astron. Soc. 420, 2851--2858.

\item 
Moskovitz, N., Gaidos, E., 2011.
Differentiation of planetesimals and the thermal consequences of melt migration.
Meteorit. Planet. Sci. 46, 903--918.

\item
Neumann, W., Breuer, D., Spohn, T., 2012.
Differentiation and core formation in accreting planetesimals.
Astron. Astrophys. 543, A141.

\item
Nimmo, F., Agnor, C.B., 2006. 
Isotopic outcomes of N-body accretion simulations: Constraints on equilibration processes during large impacts from Hf/W observations. 
Earth Planet. Sci. Lett. 243, 26--43.

\item 
Nimmo, F., Kleine, T., 2007.
How rapidly did Mars accrete? Uncertainties in the Hf-W timing of core formation.
Icarus 191. 497--504.

\item 
Nimmo, F., Hart, S.D., Korycansky, D.G., Agnor, C.B., 2008.
Implications of an impact origin for the martian hemispheric dichotomy.
Nature 453, 1220--1223.

\item 
Nimmo, F., O'Brien, D.B., Kleine, T., 2010.
Tungsten isotopic evolution during late-stage accretion: Constraints on Earth-Moon equilibration.
Earth Planet. Sci. Lett. 292, 363--370.

\item
O'Brien, D.P., Morbidelli, A., Levison, H.F. 2006.
Terrestrial planet formation with strong dynamical friction.
Icarus 184, 39--58.    
  
\item 
Ormel, C.W., Dullemond, C.P., Spaans, M., 2010.  
A new condition for the transition from runaway to oligarchic growth.
Astrophys. J. Lett. 714, L103--L107.
    
\item
Ricard, Y., \v{S}r\'{a}mek, O., Dubuffet, F., 2009. 
A multi-phase model of runaway core-mantle segregation in planetary embryos. 
Earth Planet. Sci. Lett. 284, 144--150.
  
\item 
Righter, K. and  Drake, M.J., 1996.
Core formation in Earth's Moon, Mars and Vesta. 
Icarus 124, 513--529.

\item
Righter K., Chabot, N.L., 2011.
Moderately and slightly siderophile element constraints on the depth and extent of melting in early Mars.
Meteorit. Planet. Sci. 46,  157--176.

\item
Rubie, D.C., Nimmo, F., Melosh, H.J., 2007. 
Formation of Earth's core. In: Gerald, S.,
Stevenson, D.J. (Eds.), Evolution of the Earth, Treatise on Geophysics, vol. 9.
Elsevier Scientific Publishing Company, New York, pp. 51--90 (Chapter 9.03). 
  
\item  
Rubie, D.C., Melosh, H.J., Reid, J.E., Liebske, C., Righter, K., 2003. 
Mechanisms of metal-silicate equilibration in the terrestrial magma ocean. 
Earth Planet. Sci. Lett. 205, 239--255.

\item 
Rudge J.F., Kleine, T., Bourdon, B., 2010.
Broad bounds on Earth's accretion and core formation constrained by geochemical models.
Nature Geosci. 3, 439--443.

\item 
Samuel, H., 2012.
A re-evaluation of metal diapir breakup and equilibration in terrestrial magma oceans.
Earth Planet. Sci. Lett. 313--314, 105--114.
  
\item 
Sasaki, T.,  Abe Y., 2007.
Rayleigh-Taylor instability after giant impacts: imperfect equilibration of the Hf-W system and its effects on the core formation age. 
Earths Planets Space 59, 1035--1045.  
  
 \item
\v{S}r\'{a}mek, O., Milelli, L., Ricard, Y., Labrosse, S., 2012.
Thermal evolution and differentiation of planetesimals and planetary embryos.
Icarus 217, 339--354.  

\item 
Tanaka, H., Ida, S., 1997.
Distribution of planetesimals around a protoplanet in the nebula gas. II: Numerical simulations.
Icarus 125,  302--316.
  
\item 
Tonks, W.B., Melosh, H.J., 1992.
Core formation by giant impacts. 
Icarus 100, 326--346. 
   
\item 
Tonks, W.B., Melosh, H.J., 1993.
Magma ocean formation due to  giant impacts. 
J. Geophys. Res. 98, 5319--5333.
   
\item 
Touboul, M., Kleine, T., Bourdon, B., Palme, H., Wieler, R. 2007.
Late formation and prolonged differentiation of the Moon inferred from W isotopes in lunar metals.
Nature 450, 1206--1209.
  
\item
Walsh, K.J., Morbidelli, A., Raymond, S., O'Brien, D.P., Mandell, A.M., 2011.
A low mass for Mars from Jupiter's early gas-driven migration.  
Nature 475, 206--209.  
  
\item 
Wetherill, G.W., Stewart, G.R., 1993.
Formation of planetary embryos: Effects of fragmentation, low relative velocity, and independent variation of eccentricity and inclination.
Icarus 106, 190--209.
     
\item     
Wilhelms, D.E., Squyres, S.W., 1984. 
The martian dichotomy may be due to a giant impact. 
Nature 309, 138--140.

\end{description}

\clearpage

\begin{figure}

\begin{center}
\includegraphics[width=.75\textwidth]{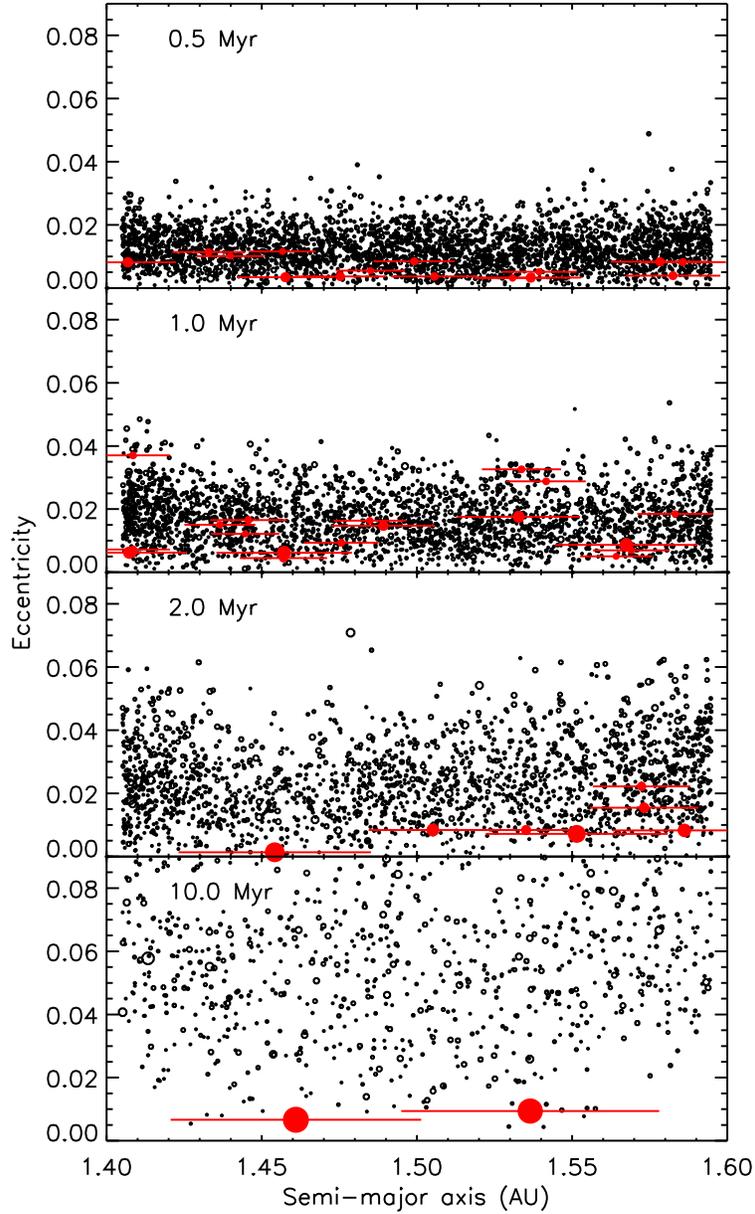}
\end{center}

\caption{Snapshots in the $a$-$e$ plane for  the $N$-body simulation with nebular gas (simulation A).
The circles are proportional to the radii of planetesimals and embryos.
Embryos are displayed as red filled circles with horizontal branches with a side length corresponding to $5 r_{\rm H}$.
The total numbers of bodies are 3,397, 2,721, 2,012, and 954 from the top panel to the bottom. }

\end{figure}

\clearpage

\begin{figure}

\begin{center}
\includegraphics[width=.8\textwidth]{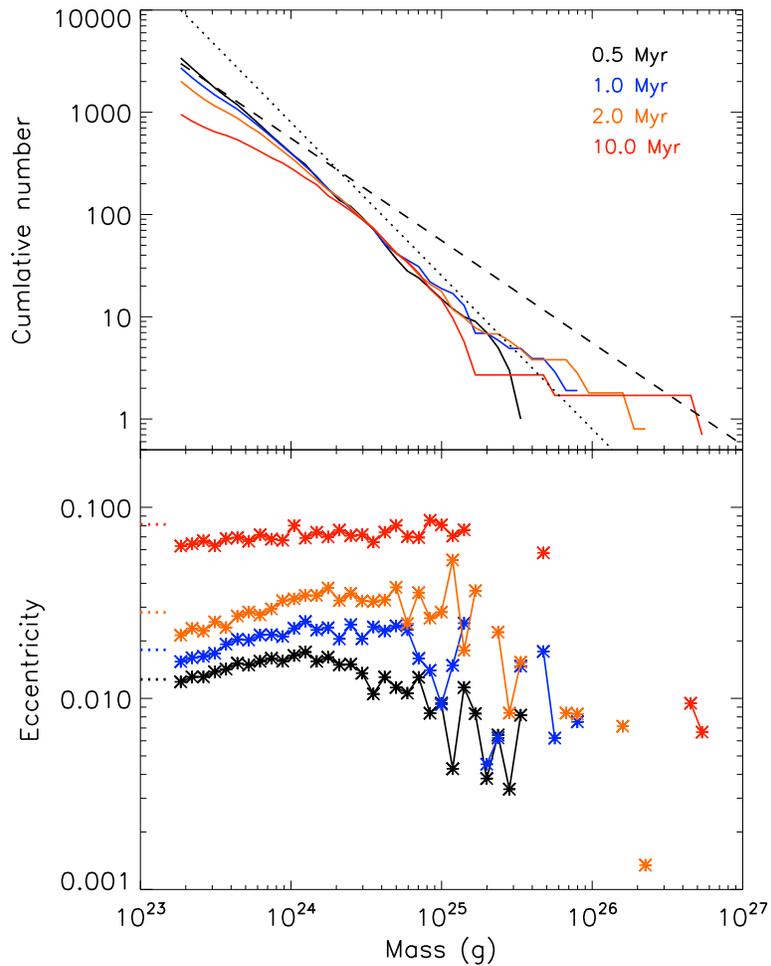}
\end{center}

\caption{ Cumulative number and orbital eccentricity as a function of mass for simulation~A.
The distributions at the four times from Fig.~1 are plotted. The dashed and dotted lines in the upper panel 
are slopes with $q = -2.0$ and $-2.5$. The dashed lines in the lower panel show the analytically estimated values for
the smallest planetesimal. }

\end{figure}

\clearpage

\begin{figure}

\begin{center}
\includegraphics[width=.8\textwidth]{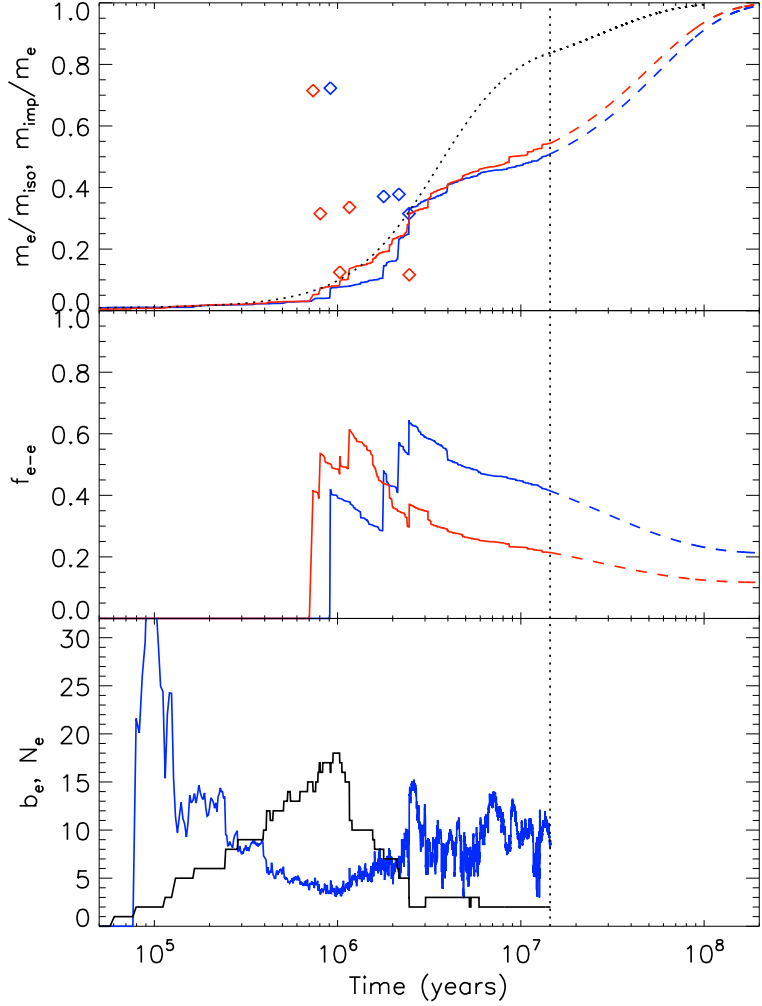}

\end{center}

\caption{(Top) The time evolution of the masses of the two surviving embryos relative to the isolation mass ($10^{27}$ g) for simulation~A.
Diamonds indicate the impactor-to-embryo mass ratios ($\le 1$) for embryo-embryo collisions. 
The dotted curve is the embryo mass derived from the analytic model. 
The vertical dotted line indicates the time at the end of simulation.
In the top and middle panels, the red (blue) color is used for the inner (outer) surviving embryo,
and the extrapolated curves until the end of compete sweep up of planetesimals are shown by dashed curves.
(Middle) Mass fractions of the embryos obtained by embryo-embryo collisions.
(Bottom) Number of embryos $N_{\rm e}$ (black line) and the mean separation $b_{\rm e}$ (blue line) of neighboring embryos normalized by the mutual Hill radius.}

\end{figure}

\clearpage

\begin{figure}

\begin{center}
\includegraphics[width=.75\textwidth]{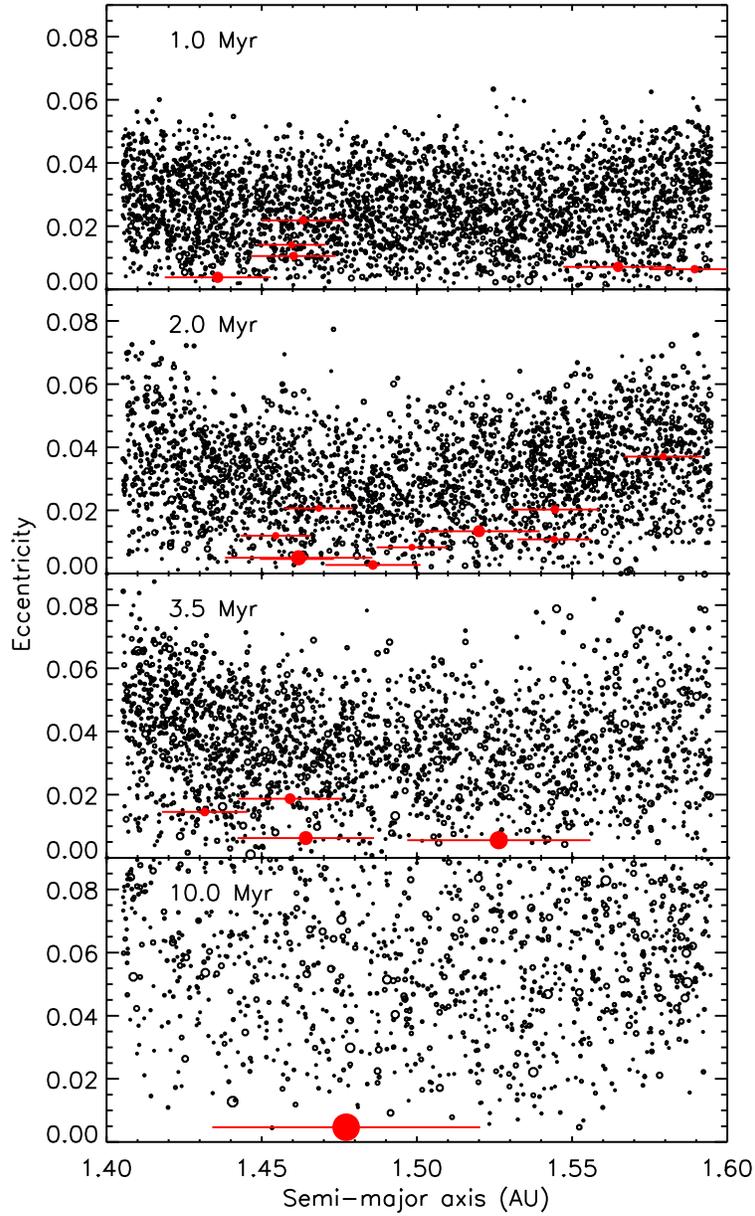}

\end{center}

\caption{Same as Fig.~1, but for the simulation without gas (simulation B). There is only one surviving embryo in this case.
The total numbers of bodies are 3,415, 2,874, 2,333, and 1,392 from the top panel to the bottom. }

\end{figure}

\clearpage

\begin{figure}

\begin{center}
\includegraphics[width=.8\textwidth]{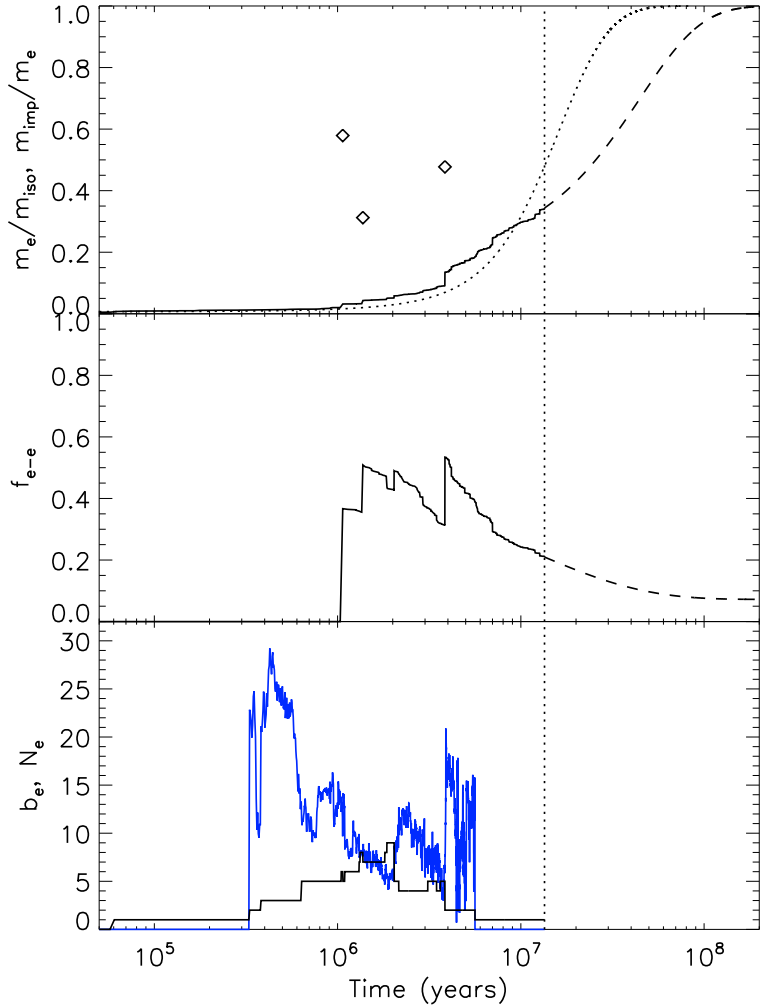}

\end{center}

\caption{Same as Fig.~3, but for simulation B.
The isolation mass is set to be $2\times 10^{27}$ g. 
For the calculation of the analytic growth curve (dotted curve) in the top panel,
the velocity dispersion is assumed to be half the escape velocity of the largest embryo. }

\end{figure}
\clearpage

\begin{figure}

\begin{center}
\includegraphics[width=.8\textwidth]{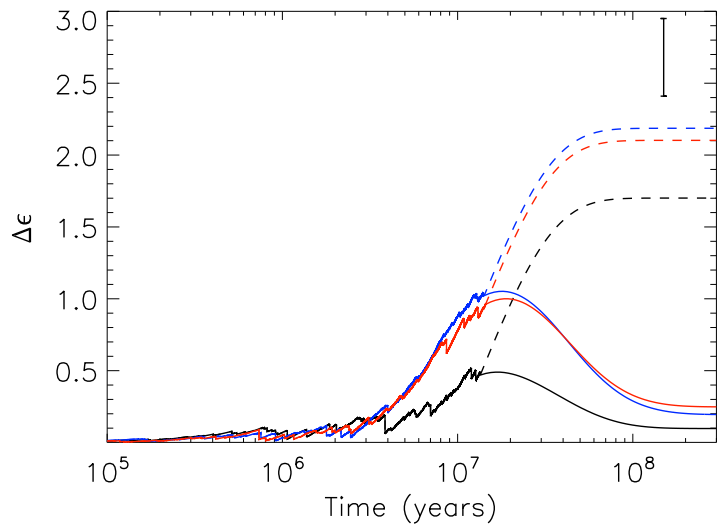}

\end{center}

\caption{Time evolution of $\Delta \epsilon$ for the surviving embryos for the case of perfect equilibration between 
the target's mantle and impactors ($k_{\rm tm} = k_{\rm ic} = 1$). 
The red and blue curves are for the inner and outer embryos from simulation A
whereas the black curve is for the embryo from simulation B.
 The solid curves are cases where the growth curves of the embryos are extrapolated to complete sweep up of planetesimals
 while the dashed curves are cases where accretion is stopped at the end of $N$-body simulations ($t \sim 14$ Myr).
The error bar at the upper right represents $\Delta \epsilon$ for present-day Mars.}

\end{figure}

\clearpage

\begin{figure}

\begin{center}
\includegraphics[width=.8\textwidth]{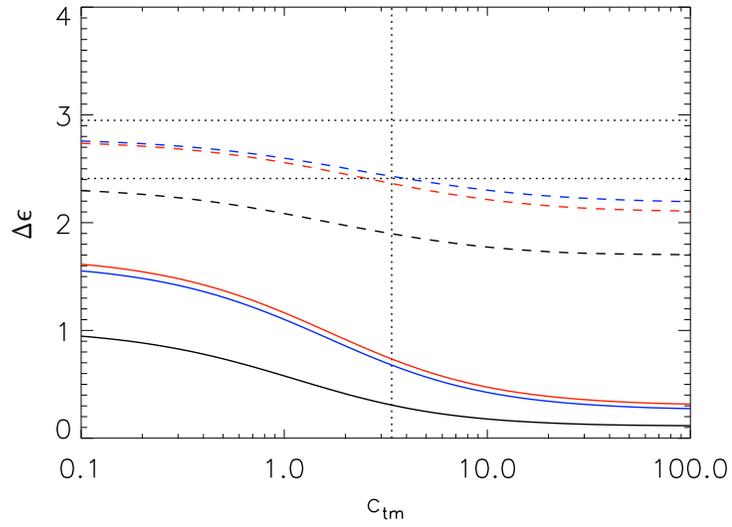}

\end{center}

\caption{Present-day values of $\Delta \epsilon$ for the surviving embryos for the case with 
partial equilibration of the target's mantle ($k_{\rm tm} \le 1.0$ and $k_{\rm ic} = 1.0$). 
Eq.~(\ref{eq:ktm}) is used for modeling of $k_{\rm tm}$. 
The meanings of curves are the same as in Fig.~6.
The region between the two horizontal dashed lines represents the range of $\Delta \epsilon$ for present-day Mars 
while the vertical dashed line represents $c_{\rm tm} = f^{\rm Hf/W} = 3.38$.}
 
\end{figure}

\clearpage

\begin{figure}

\begin{center}
\includegraphics[width=.8\textwidth]{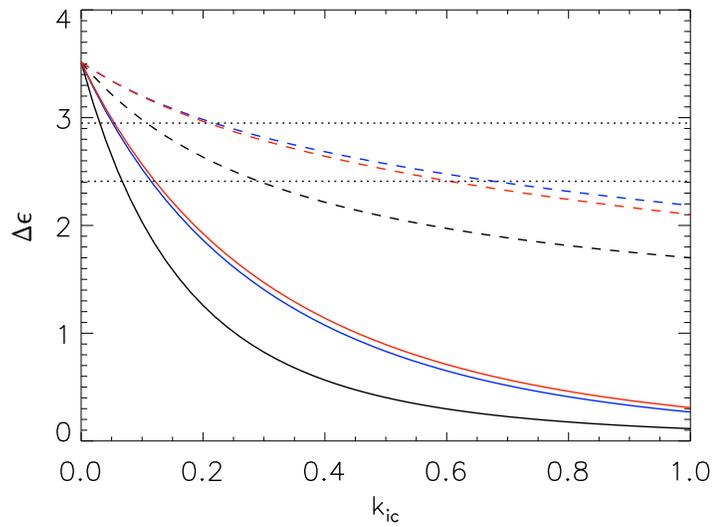}

\end{center}

\caption{Present-day values of $\Delta \epsilon$ for the surviving embryos 
as a function of $k_{\rm ic}$ ($k_{\rm tm} = 1.0$). 
The meanings of the curves and the lines are the same as those in Fig.~7.}

\end{figure}

\clearpage

\begin{figure}

\begin{center}
\includegraphics[width=.8\textwidth]{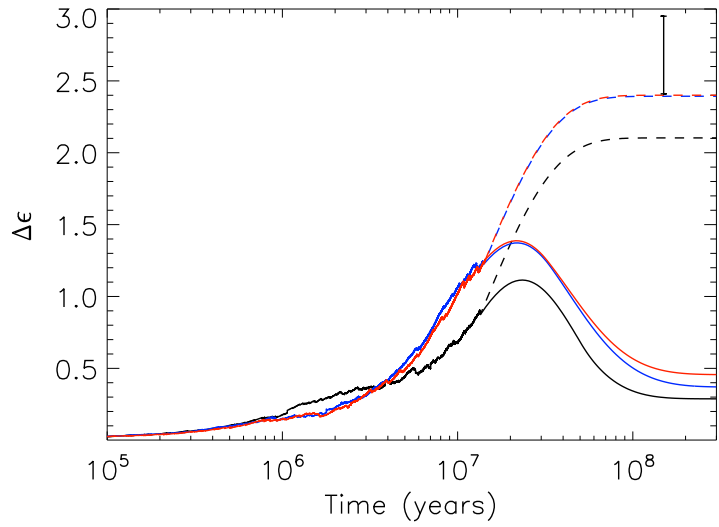}
\end{center}

\caption{Same as Fig.~6 but for the case with partial equilibration of
both the target's mantle and the impactors' cores.
Eq.~(\ref{eq:ktm}) with $c_{\rm tm} = 10$ is used for the modeling of 
$k_{\rm tm}$ whereas Eq.~(\ref{eq:kic}) with $c_{\rm ic} = 0.01$ 
is used for the modeling of $k_{\rm ic}$. }

\end{figure}

\clearpage

\begin{figure}

\begin{center}
\includegraphics[width=.9\textwidth]{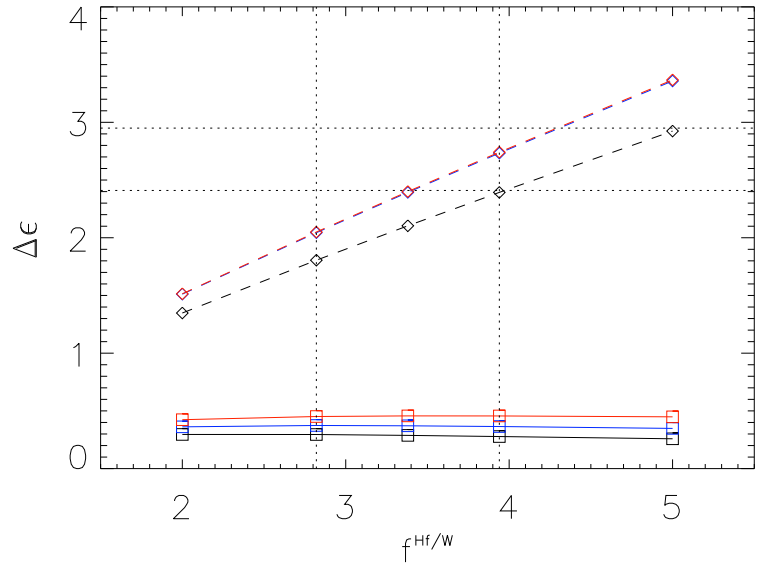}

\end{center}

\caption{Present-day values of $\Delta \epsilon$ for the surviving embryos 
as a function of $f^{\rm Hf/W}$. The partial equilibration 
model from Fig.~9 is used.
The meanings of the solid and dashed lines, and the horizontal dotted lines are the same as those in Fig.~7.
The region between the two vertical dotted lines represents the range of $f^{\rm Hf/W} $ estimated by Dauphas and Pourmand (2011).}

\end{figure}

\end{document}